\documentstyle[11pt,epsfig,amssymb]{article}
\title{\bf Energy conditions in $f(R)$ gravity and Brans-Dicke theories}
\author{K. Atazadeh\thanks{email: k-atazadeh@sbu.ac.ir}, A.
Khaleghi\thanks{email: ahadkhaleghi@gmail.com},\,\, H. R.
Sepangi\thanks{email: hr-sepangi@sbu.ac.ir}\,\,\ and Y.
Tavakoli\thanks{email: y-tavakoli@std.sbu.ac.ir}
\\
{\small Department of Physics, Shahid Beheshti University, Evin,
Tehran 19839, Iran}}
\begin{document}
\maketitle

\begin{abstract}
The equivalence between $f(R)$ gravity and scalar-tensor theories is
invoked to study the null, strong, weak and dominant energy
conditions in Brans-Dicke theory. We consider the validity of the
energy conditions in Brans-Dicke theory by invoking the energy
conditions derived from a generic $f(R)$ theory. The parameters
involved are shown to be consistent with an accelerated expanding
universe.
\end{abstract}
\vspace{2cm}

\section{Introduction}
Recent observations have revealed that the present state of the
universe is undergoing an accelerated expansion \cite{1}. There are,
in general, a number of different approaches towards explaining this
acceleration. One such approach utilizes what is known as $f(R)$
modification of gravity which, in effect is equivalent to
Brans-Dicke (BD) type theories. The assumption of the existence of
dark energy is another approach often used in this respect. In all
the above theories, energy conditions impose stringent constraints
whose validity should be studied in the light of their ability to
explain the observational data. From a theoretical viewpoint, energy
conditions in their various forms, namely strong energy condition
(SEC), weak energy condition (WEC), dominant energy condition (DEC),
and null energy condition (NEC) have been used in different contexts
to derive general results that would hold for a variety of
situations \cite{2}. For example, the Hawking-Penrose singularity
theorems invoke the WEC and SEC \cite{3}, whereas the proof of the
second law of black hole thermodynamics requires the NEC \cite{4}.
Another example comes from cosmology \cite{5} where energy
conditions are studied by using red shifts.

The equivalence between BD theories and $f(R)$ gravity is a subject
that has been studied by various authors, as an example see
\cite{6}. The study of energy conditions may thus benefit from such
equivalence in that knowing the energy conditions in one theory
would point to the energy conditions in the other. For example, in
the BD theory case $V(\phi)=V_{0}\phi^{2}$ \cite{berto} where
$V_{0}$ cannot be found easily, one can use the equivalent $f(R)$
theory to facilitate the calculation of $V_{0}$ which, could then be
used in the BD theory.

Even though the goal of this paper is to study energy conditions
in modified theories of gravity and consequently in Brans-Dicke
theory, much of the techniques will be borrowed from the analysis
of energy conditions in Einstein's gravity. Therefore, we shall
briefly review the derivation of energy conditions in general
relativity. For a pedagogical review, see for example \cite{3}.

\section{Energy conditions in general relativity}
Let $\upsilon^{\mu}$ be a tangent vector to a congruence of time
like geodesics. For a hypersurface orthogonal congruence,
Raychaudhuri's equation reads
\begin{equation}\label{eq1}
\frac{d\theta}{d\tau}=-\frac{1}{3}
\theta^{2}-\sigma_{\mu\nu}\sigma^{\mu\nu}-R_{\mu\nu}\upsilon^{\mu}\upsilon^{\nu},
\end{equation}
where $\theta$ and $\sigma^{\mu\nu}$ are the expansion and sheer of
two nearby tangent vectors, respectively. In Einstein's gravity, SEC
and the ensuing singularities theorem follow from requiring that
\begin{equation}\label{eq2}
R_{\mu\nu}\upsilon^{\mu}\upsilon^{\nu}=8\pi
G(T_{\mu\nu}-\frac{1}{2}g_{\mu\nu}T)\upsilon^{\mu}\upsilon^{\nu}\geq
0,
\end{equation}
for all time-like $\upsilon^{\mu}$ for which
$\frac{d\theta}{d\tau}<0$. A pair of nearby time-like geodesic
vectors converge and will eventually intersect. The stress-energy
tensor at each point $p \in M= \Re^{4}$ obeys the inequality
$T_{\mu\nu}\upsilon^{\mu}\upsilon^{\nu}\geq 0$ for any time like
vector $\upsilon^{\mu}\in \textbf{T}_{p}$ of an observer whose
world-line at $p$ has the unit tangent vector $\textbf{v}$ and the
local energy density appears to be $T_{\mu\nu}$. This assumption
is thus equivalent to the energy density being non-negative as
measured by any observer which, of course, is physically
reasonable.
\section{Energy condition in $f(R)$ gravity}
In this section we use the metric formalism in $f(R)$ gravity and
derive the strong, weak and dominant energy conditions for a
general form of $f(R)$. In doing so we will follow the formalism
recently developed in \cite{7}. We take the
Freedman-Robertson-Walker (FRW) metric to study the cosmological
implications of the models studied here.

The action for $f(R)$ gravity is \cite{8}
\begin{equation}\label{eq3}
S=\frac{1}{2\kappa}\int d^{4}x \sqrt{-g}f(R)+{\cal S}_{m},
\end{equation}
where we have set $\kappa=8\pi G=1$. The field equations resulting
from this action in the metric approach, assuming the connections
are that of the Levi-Civita, are given by
\begin{equation}\label{eq4}
G_{\mu\nu}=R_{\mu\nu}
-\frac{1}{2}g_{\mu\nu}R=T_{\mu\nu}^{c}+\tilde{T}_{\mu\nu}^{m},
\end{equation}
where $\tilde{T}_{\mu\nu}^{m}=\frac{T_{\mu\nu}^{m}}{f'(R)}$
represents the energy-momentum tensor of ordinary matter
considered as perfect fluid given by
\begin{equation}\label{eq5}
T_{\mu\nu}^{m}=(\rho+p)u_{\mu}u_{\nu}+pg_{\mu\nu},
\end{equation}
and $T_{\mu\nu}^{c} $ is the stress energy tensor of the
gravitational fluid
\begin{equation}\label{eq6}
T_{\mu\nu}^{c}=\frac{1}{f'(R)}\left[\frac{1}{2}g_{\mu\nu}(f(R)-Rf'(R))
+(g_{\alpha\mu}g_{\beta\nu}-g_{\mu\nu}g_{\alpha\beta})\nabla^{\alpha}\nabla^{\beta}f'(R)\right],
\end{equation}
where a prime represents differentiation with respect to $R$. The
field equation (\ref{eq4}) now reads
\begin{equation}\label{eq7}
f'(R)R_{\mu\nu}-\frac{1}{2}f(R)g_{\mu\nu}-(\nabla_{\mu}\nabla_{\nu}-g_{\mu\nu}\Box)f'(R)=T_{\mu\nu}^{m}.
\end{equation}
Contracting the above equation we obtain
\begin{equation}\label{eq8}
f'(R)R-2f(R)+3\Box f'(R)=T^{m}.
\end{equation}

Now, let us briefly review the energy conditions in $ f(R)$
gravity. We begin by defining an effective stress-energy tensor
using equation (\ref{eq7}) as follows
\begin{equation}\label{eq9}
T_{\mu \nu }^{\rm
eff}=\frac{1}{f'(R)}\left[T_{\mu\nu}^{m}+\frac{1}{2}\left(f(R)-Rf'(R)\right)g_{\mu\nu}+
(\nabla_{\mu}\nabla_{\nu}-g_{\mu\nu}\Box)f'(R)\right],
\end{equation}
with
\begin{equation}\label{eq10}
T^{\rm eff}=\frac{1}{f'(R)}\left[T^{m}+2(f(R)-Rf'(R))-3\Box
f'(R)\right].
\end{equation}
Therefore, we can write $R_{\mu\nu}$ in the terms of an effective
stress-energy tensor and its trace, that is
\begin{equation}\label{eq11}
R_{\mu\nu}=\kappa\left(T^{\rm
eff}_{\mu\nu}-\frac{1}{2}g_{\mu\nu}T^{\rm eff}\right).
\end{equation}
Using the spatially flat $(k=0)$ FRW metric as
\begin{equation}\label{eq12}
ds^{2}=-dt^{2}+a(t)^{2}d\textbf{x}^{2},
\end{equation}
the effective energy density and pressure are given by
\begin{equation}\label{eq13}
\rho^{\rm eff}=\frac{1}{f'(R)}\left[\rho-
\frac{1}{2}(f(R)-Rf'(R))-3H\dot{R}f''(R)\right],
\end{equation}
and
\begin{equation}\label{eq14}
p^{\rm eff}=\frac{1}{f'(R)}\left[p+
\frac{1}{2}(f(R)-Rf'(R))+(2\dot{R}H+\ddot{R})f''(R)+\dot{R}^{2}f'''(R)\right],
\end{equation}
where $\dot{R}=dR/dt$ and $H(t)=\frac{\dot{a}(t)}{a(t)}$ is the
Hubble parameter. Now, using these equations, we can write the NEC
and SEC, given by $\rho^{\rm eff}+p^{\rm eff}\geq 0$ and $\rho^{\rm
eff}+3p^{\rm eff}\geq 0$ respectively as
\begin{equation}\label{eq15}
\rho+p+(\ddot{R}-\dot{R}H)f''(R)+\dot{R}^{2}f'''(R)\geq
0
\end{equation}
and
\begin{equation}\label{eq16}
\rho+3p+(f(R)-Rf'(R))+3(\ddot{R}+\dot{R}H)f''(R)+3
\dot{R}^{2}f'''(R)\geq 0.
\end{equation}

To compare our results here with that of general relativity for a
given $f(R)$, we use the FRW metric which, for WEC ($\rho^{\rm
eff}\geq 0$) leads to
\begin{equation}\label{eq17}
\rho- \frac{1}{2}(f(R)-Rf'(R))-3H\dot{R}f''(R)\geq 0.
\end{equation}
For DEC ($\rho^{\rm eff}-p^{\rm eff}\geq 0$), we find
\begin{equation}\label{eq18}
\rho-p-(5\dot{R}H+\ddot{R})f''(R)-\dot{R}^{2}f'''(R)-(f(R)-Rf'(R))\geq 0.
\end{equation}
\section{Energy conditions in Brans-Dicke theory}
Let us now investigate a non-minimally coupled self interacting
scalar-tensor field theory such as the Brans-Dicke (BD) theory and
find the various energy conditions for this type of modified
gravity. In the context of BD theory \cite{9} with a self
interacting potential and a matter field, the action is given by
\begin{equation}\label{eq19}
S=\frac{1}{2\kappa}\int d^{4}x \sqrt{-g}\left[\phi
R-\frac{\omega}{\phi}\nabla^{\alpha}\phi\nabla_{\alpha}\phi-V(\phi)\right]+{\cal
S}_{m},
\end{equation}
where $\omega$ is the usual BD parameter and we have chosen units
such that $8\pi G=c=1$. The gravitational field equations can be
derived from action (\ref{eq18}) by varying the action with
respect to the metric
\begin{equation}\label{eq20}
G_{\mu\nu}=\frac{T_{\mu\nu}^{m}}{\phi}+\frac{\omega}{\phi^{2}}\left(\nabla_{\mu}\phi
\nabla_{\nu}\phi-\frac{1}{2}g_{\mu\nu}\nabla^{\alpha}\phi\nabla_{\alpha}\phi\right)+\frac{1}{\phi}
\left(\nabla_{\mu}\nabla_{\nu}\phi-g_{\mu\nu}\Box\phi\right)
-g_{\mu\nu}\frac{V(\phi)}{2\phi},
\end{equation}
where $T_{\mu\nu}^{m}$ is the stress-energy tensor of the normal
matter as expressed in equation (\ref{eq5}). Variation of action
(\ref{eq19}) with respect to  $\phi$ gives
\begin{equation}\label{eq21}
\Box\phi=\frac{T^{m}}{2\omega+3}+\frac{1}{2\omega+3}\left[\phi\frac{dV(\phi)}{d\phi}-2V(\phi)\right],
\end{equation}
where the expression
$\Box\phi=g^{\mu\nu}\nabla_{\mu}\nabla_{\nu}\phi$, using
(\ref{eq12}) is given by
\begin{equation}\label{eq22}
\Box\phi=-(\ddot{\phi}+3H\dot{\phi}).
\end{equation}
Using the equation of motion, we can write $G_{\mu\nu}=\kappa'(
T_{\mu\nu}^{\rm m}+T_{\mu\nu}^{\phi})$   where
$\kappa'=\frac{1}{\phi}$. The stress-energy tensor and its trace
for the BD theory may now be calculated with the result
\begin{equation}\label{eq23}
T^{\phi}_{\mu\nu}=\frac{\omega}{\phi}[\nabla_{\mu}\phi\nabla_{\nu}\phi-\frac{1}{2}g_{\mu\nu}\nabla^{\alpha}
\phi\nabla_{\alpha}\phi]+[\nabla_{\mu}\nabla_{\nu}\phi-g_{\mu\nu}\Box\phi]
-g_{\mu\nu}\frac{V(\phi)}{2},
\end{equation}
and
\begin{equation}\label{eq24}
T^{\phi}=\frac{\omega}{\phi}[(\nabla\phi)^{2}-2\nabla^{\alpha}
\phi\nabla_{\alpha}\phi]-3\Box\phi-2V(\phi).
\end{equation}
Comparison of these equations with equation (\ref{eq2}) leads to a
similar equation for the SEC,
\begin{equation}\label{eq25}
R_{\mu\nu}\upsilon^{\mu}\upsilon^{\nu}=\kappa'\left[ (T^{\rm
m}_{\mu\nu}+T_{\mu\nu}^{\phi})-\frac{1}{2}g_{\mu\nu}(T^{\rm
m}+T^{\phi})\right]\geq 0.
\end{equation}
We can now write the relations for NEC and SEC analogously as
$(\rho+\rho^{\phi}+ p+p^{\phi})\geq 0 $ and $(\rho+\rho^{\phi}+
3(p+p^{\phi}))\geq 0$ respectively \cite{3} so that we may first
derive $\rho^{\phi}$ and $p^{\phi}$ for the spatially flat FRW
metric as follows
\begin{equation}\label{eq25}
\rho^{\phi}=\frac{\omega}{2\phi}\dot{\phi}^{2}-3H\dot{\phi}+\frac{V(\phi)}{2},
\end{equation}
and
\begin{equation}\label{eq26}
p^{\phi}=\frac{\omega}{2\phi}\dot{\phi}^{2}+(2H\dot{\phi}+\ddot{\phi})-\frac{V(\phi)}{2}.
\end{equation}

Thus, the NEC and SEC for the BD theory are given by
\begin{equation}\label{eq28}
\rho+p+\frac{\omega}{\phi}\dot{\phi}^{2}+(\ddot{\phi}
-H\dot{\phi})\geq 0,
\end{equation}
and
\begin{equation}\label{eq29}
\rho+3p+\frac{2\omega}{\phi}\dot{\phi}^{2}+3(\ddot{\phi}+H\dot{\phi})-V(\phi)\geq
0.
\end{equation}
Now, following and expanding on the GR approach to include WEC and
DEC, as has been employed in $f(R)$ gravity theories, we may
obtain similar equations in the BD theory. Therefore, the WEC and
DEC in the BD theory are respectively given by
\begin{equation}\label{eq30}
\rho+\frac{\omega}{2\phi}\dot{\phi}^{2}-3H\dot{\phi}+\frac{V(\phi)}{2}\geq
0
\end{equation}
and
\begin{equation}\label{eq31}
\rho-p-(\ddot{\phi}+5H\dot{\phi})+V(\phi)\geq 0.
\end{equation}
\section{Equivalence of the energy conditions in $f(R)$ gravity and Brans-Dicke theory}
Considering action (\ref{eq3}) within the context of the metric
formulation of $f(R)$ gravity, one can introduce a new field $\chi$
and write a dynamically equivalent action \cite{6}
\begin{equation}\label{eq32}
S=\frac{1}{2\kappa}\int d^{4}x
\sqrt{-g}\left[f(\chi)+f'(\chi)(R-\chi)\right]+{\cal
S}_{m}(g_{\mu\nu},\psi).
\end{equation}
Variation with respect to $\chi$ leads to equation $\chi=R$
provided $f''(R)\neq0$, which reproduces action (\ref{eq3}).
Redefining the field  $\chi$ by $\phi=f'(R)$ and setting
\begin{equation}\label{eq33}
V(\phi)=\phi\chi(\phi)-f(\chi(\phi)),
\end{equation}
the action takes the form
\begin{equation}\label{eq34}
S=\frac{1}{2\kappa}\int d^{4}x \sqrt{-g}(\phi R-V(\phi))+{\cal
S}_{m}(g_{\mu\nu},\psi).
\end{equation}
Comparison with action (\ref{eq19}) reveals that this is the action
of a BD theory with $\omega=0$. Therefore, metric $f(R)$ theories,
as has been observed long ago, are fully equivalent to a class of BD
theories with a vanishing kinetic term \cite{6}. Now, taking
$\chi=R$ and $\phi=f'(R)$ we have
\begin{equation}\label{eq35}
\dot{\phi}=\dot{R}f''(R)~~~~~~\mbox{and}~~~~~~~~~
\ddot{\phi}=\ddot{R}f''(R)+\dot{R}^{2}f'''(R).
\end{equation}
Substituting these relations into equations for NEC, SEC, WEC and
DEC in BD theory, namely equations (\ref{eq28}), (\ref{eq29}),
(\ref{eq30}) and (\ref{eq31}) with $\omega=0$ , one can easily
derive respectively NEC, SEC, WEC and DEC for the $f(R)$
modification of gravity, that is  equations (\ref{eq15}),
(\ref{eq16}), (\ref{eq17}) and (\ref{eq18}).
\subsection{Examples}
To see how equation (\ref{eq17}) can be used to put constraints on a
given $f(R)$ and equivalently on the BD potential, let us examine
two examples. First, we consider $f(R)$ as having a general
power-law form, given by
\begin{equation}\label{eq36}
f(R)=\sigma R^{n}.
\end{equation}
Let us now concentrate on the vacuum sector {\it i.e.} $\rho=p=0$.
Substituting in equation (\ref{eq17}) we have the following
condition for WEC
\begin{equation}\label{eq37}
\sigma (n-1)(1-nA)\geq0,
\end{equation}
where $A=\frac{j_{0}-q_{_{0}}-2}{(1-q_{_{0}})^{2}}$ and the
deceleration $(q_{_{0}})$, jerk $(j_{_{0}})$ and snap $(s_{_{0}})$
parameters for the present-day values are defined in \cite{7}. I
what follows, we examine two values of the exponent $n$, namely
$n=-1$ and $n=2$
which satisfy the inequality (\ref{eq37}).\vspace{2mm}\noindent\\
\textbf{Case I}: $n=-1$ \vspace{1mm}\noindent\\
Takeing $q_{_{0}}\sim -0.81$ and $j_{_{0}}\sim 2.16$ given in
\cite{7}, so that $A\sim 0.29$. Equation (\ref{eq37}) for $n=-1$
reduces to
$$-2.58\sigma \geq0.$$  This relation is satisfied for $\sigma<0$
which in turn satisfies equation (\ref{eq37}) with $n=-1$. The
potential in BD theory with $\omega=0$ for $V(\phi)=V_{0}\phi^{m}$,
corresponding to $f(R)=\sigma R^{n}$ can thus be obtained simply as
$$V_{0}=\frac{(n-1)\sigma}{(n\sigma)^{m}}.$$ where
$m=\frac{n}{n-1}$, so that for $n=-1$ the corresponding BD potential
is $V(\phi)=-2\sqrt{-\sigma}\phi^{\frac{1}{2}}$.\vspace{2mm}\noindent\\
\textbf{Case II}: $n=2$\vspace{1mm}\noindent\\
As a second case we consider $n=2$, so that $f(R)=\sigma R^{2}$,
leading to a WEC given by $-0.06 \sigma \geq0$ which requires
$\sigma <0$. The corresponding BD potential is
$V(\phi)=\frac{1}{4\sigma} \phi^{2}$ with $\sigma$ being negative.
One can therefore come to the conclusion that  $V_{0}$ must also
have a negative value.
\section{Redshift and energy conditions}
Let us now take the potential $V(\phi)=V_{0}\phi^{m}$ in BD theory
with $\omega=0$ and use the following power-low ans\"{a}tze
\begin{equation}\label{eq38}
\phi=\phi_{0}\left(\frac{t}{t_{0}}\right)^{\beta} ~~~~~~~~~
\mbox{and} ~~~~~~~~
a(t)=a_{0}\left(\frac{t}{t_{0}}\right)^{\alpha}.
\end{equation}
Inserting  these relations into equation (\ref{eq21}) in the
vacuum sector, one then finds that
\begin{equation}\label{eq39}
\beta =\frac{2}{1-m}.
\end{equation}
Substituting the spatially flat FRW metric (\ref{eq12}) in the field
equations (\ref{eq4}) we get
\begin{equation}\label{eq39}
3\left(\frac{\dot{a}}{a}\right)^{2}=\rho^{\rm eff},
\end{equation}
and
\begin{equation}\label{eq40}
2\frac{\ddot{a}}{a}+\left(\frac{\dot{a}}{a}\right)^{2}=-p^{\rm eff}.
\end{equation}
Now, from equations (\ref{eq38}) and (\ref{eq39}) we can write
\begin{equation}\label{eq42}
\rho^{\rm eff}=3H_{0}^{2}(1+z)^{\frac{2}{\alpha}},
\end{equation}
where we have used the relations  $\frac{a(t_{0})}{a(t)}=1+z$,
$H_{0}=\frac{\dot{a}(t_{0})}{a(t_{0})}$ and $\alpha=t_{0}H_{0}$
where $z$ is the redshift of a luminous source \cite{Coles}. From
this equation one may conclude that in the very early universe
$\alpha$ would have been small and therefore $\rho^{\rm eff}$ large.
Now, the WEC is given by
\begin{equation}\label{eq43}
3H_{0}^{2}(1+z)^{\frac{2}{\alpha}} \geq 0.
\end{equation}
This equation for the effective energy density clearly satisfies
the WEC. If we follow the same method as in section 3 or in
\cite{7} and substitute the total energy density and pressure by
$\rho^{\rm eff}$ and $p^{\rm eff}$, we will find the same relation
between the distance modulus and redshift parameter which is
studied in \cite{Coles} where the energy conditions have been
used. Now let us write the NEC, DEC and SEC with respect to the
redshift parameter as follows
\begin{equation}\label{eq44}
\frac{2H_{0}}{t_{0}}(1+z)^{\frac{2}{\alpha}} \geq 0,
\end{equation}
\begin{equation}\label{eq45}
(6 \alpha-2)\frac{H_{0}}{t_{0}}(1+z)^{\frac{2}{\alpha}} \geq 0,
\end{equation}
\begin{equation}\label{eq46}
6(1-\alpha) \frac{H_{0}}{t_{0}}(1+z)^{\frac{2}{\alpha}} \geq 0,
\end{equation}
respectively. From equation (\ref{eq44}) one can see that it is not
possible to extract more information from NEC. As far as the DEC and
SEC are concerned however, they require  $\alpha \geq \frac{1}{3}$
and $\alpha \leq 1$ respectively. The first is in agreement with the
observation that the universe is undergoing an accelerated expansion
phase, $\alpha > 1$. A glance at the last equation reveals that it
is completely in contradiction with an accelerated expanding
universe, but we know that the SEC ensures gravity to be always
attractive. Violation, as discussed in \cite{Visser}, allows for the
late time accelerated cosmic expansion as suggested by the
combination of recent astronomical observations.
\section{Energy condition in the Einstein frame}
What we have done so far in the pervious sections has been in the
so-called Jordan frame. However, it would also be instructive to
study these relations in the Einstein frame.  As is well known,
the usual procedure, going from one frame to the other, is to use
a conformal transformation. A problem then arises in that whether
the tensor representing the \emph{physical} metric structure of
space-time is the one belonging to the Jordan frame or to the
Einstein frame. However, what we are concerned with in this work
is the relation between the energy conditions in these frames and
will not deal with the question posed above. Under the conformal
transformation
\begin {equation}\label{eq51}
\tilde{g}_{\mu\nu}\rightarrow e^{\phi}g_{\mu\nu},
\end{equation}
and taking
\begin {equation}\label{eq52}
\phi=-\ln f'(R),
\end{equation}
the action (\ref{eq3}) is rewritten as
\begin {equation}\label{eq53}
S_{_{E}}=\int
d^{4}x\sqrt{-g}\left(R-\frac{3}{2}g^{\alpha\beta}\partial_{\alpha}\phi\partial_{\beta}\phi-V(\phi)\right).
\end{equation}
Here
\begin {equation}\label{eq54}
V(\phi)=\frac{R}{f(R)}-\frac{f(R)}{f'(R)^{2}}.
\end{equation}
As a result, one finds the field equations for the metric in the
form
\begin {equation}\label{eq55}
\tilde{G}_{\mu\nu}=\frac{3}{2}\tilde{\nabla}_{\mu}\phi\tilde{\nabla}_{\nu}\phi-\frac{3}{4}
\tilde{g}_{\mu\nu}g^{\alpha\beta}\tilde{\nabla}_{\alpha}\phi\tilde{\nabla}_{\beta}\phi-\frac{1}{2}V(\phi)\tilde{g}_{\mu\nu}.
\end{equation}
Also, the equation of motion for $\phi$ becomes
\begin {equation}\label{eq56}
\tilde{\square}\phi=\frac{1}{3}\frac{dV}{d\phi}.
\end{equation}
For the FRW metric we find
\begin {equation}\label{eq57}
\tilde{\square}\phi=-e^{-\phi}(\ddot{\phi}+\dot{\phi}^{2}+3H\dot{\phi}).
\end{equation}
In order to solve these equations for the case $f(R)=\sigma R^{n}$
we use the following power-low \emph{ans\"{a}tze}
\begin {equation}\label{eq58}
 \tilde{a}(\tilde{t})\propto\tilde{t}^{\beta} ~~~~~~~~~~~~\mbox{and}~~~~~~~~~~~\phi=\alpha \ln
 \tilde{t},
\end{equation}
where $\alpha$ and $\beta$ are arbitrary constants. The Hubble
parameter, $\tilde{H}$, then reads
\begin{equation}\label{eq59}
\tilde{H}=\frac{\beta}{\tilde{t}}\,\,.
\end{equation}
Now, using equations (\ref{eq52}) and (\ref{eq54}) we have
\begin {equation}\label{eq60}
 V(\phi)=\frac{n-1}{n^{2} \sigma}
 {\left(\frac{e^{-\phi}}{n\sigma}\right)}^\frac{2-n}{n-1}.
\end{equation}
Matching the exponents of $t$ in equation (\ref{eq55}) we arrive
at the following expression for $\alpha$
\begin {equation}\label{eq61}
\alpha=\frac{2(1-n)}{2n-3}.
\end{equation}
If we define $T^{\phi}_{\mu\nu}$ as
\begin {equation}\label{eq62}
T^{\phi}_{\mu\nu}=\frac{3}{2}\tilde{\nabla}_{\mu}\phi\tilde{\nabla}_{\nu}\phi-\frac{3}{4}
\tilde{g}_{\mu\nu}\tilde{g}^{\alpha\beta}\tilde{\nabla}_{\alpha}\phi\tilde{\nabla}_{\beta}\phi-\frac{1}{2}V(\phi)\tilde{g}_{\mu\nu},
\end{equation}
then we obtain  the following relations for $\rho^{\phi}$ and
$P^{\phi}$
\begin {equation}\label{eq63}
 \rho^{\phi}=\frac{3\dot{\phi}^{2}}{4}+\frac{1}{2}e^{\phi}V(\phi) ~~~~~~~\mbox{and}~~~~~~~p^{\phi}=\frac{3a^{2}\dot{\phi}^{2}}{4}-\frac{1}{2}e^{\phi}V(\phi)a^{2}.
 \end{equation}
Now let us write the WEC, SEC, DEC, and NEC
\begin {equation}\label{eq64}
\frac{3\dot{\phi}^{2}}{2}+e^{\phi}V(\phi)\geq 0,
\end{equation}
\begin {equation}\label{eq65}
\frac{3(1+3a^{2})\dot{\phi}^{2}}{2}+e^{\phi}V(\phi)(1-3a^{2})\geq 0,
\end{equation}
\begin {equation}\label{eq66}
\frac{3(1-a^{2})\dot{\phi}^{2}}{2}+e^{\phi}V(\phi)(1+a^{2})\geq 0,
\end{equation}
\begin {equation}\label{eq67}
\frac{3(1+a^{2})\dot{\phi}^{2}}{2}+e^{\phi}V(\phi)(1-a^{2})\geq 0.
\end{equation}
Using our \emph{ans\"{a}tze} (\ref{eq58}), we find the following
expression for the WEC
\begin {equation}\label{eq68}
\frac{6(1-n)^{2}}{(2n-3)^{2}}+\left(\frac{n-1}{n^{2}\sigma}\right)\left(\frac{1}{n\sigma}\right)^{\frac{2-n}{n-1}}\geq
0.
\end{equation}
As we can see in  equation (\ref{eq68}), the WEC always holds.
However, in the case that either $n<0$ or $\sigma<0$, we have a
constraint on $n$, namely that $n$ must be even. For the other
energy conditions at late times, $a(t)\longrightarrow\infty$, we
have
\begin{equation}\label{eq69}
6\left(\frac{1-n}{2n-3}\right)^{2}-\left(\frac{n-1}{n^{2}\sigma}\right)\left(\frac{1}{n\sigma}\right)^{\frac{2-n}{n-1}}\geq0~~~~~~~\mbox{SEC
and NEC},
\end{equation}
\begin {equation}\label{eq70}
6\left(\frac{1-n}{2n-3}\right)^{2}-
\left(\frac{n-1}{n^{2}\sigma}\right)\left(\frac{1}{n\sigma}\right)^{\frac{2-n}{n-1}}\leq0
~~~~~~~~~~\mbox{DEC}.
\end{equation}
Equation (\ref{eq69}) tells us that there is no guarantee that it
remains positive. As can be seen, when the SEC and NEC hold, the DEC
does not. For these equations we also have a constraint on $n$, that
is, when either $n<0$ or $\sigma<0$, then $n$ must be even. In the
case of $n<0$ another constraint appears which is important for both
equations. There are ranges of $n$ for which either SEC or NEC do
not hold. Figure 1 shows this situation. Also, for the case $n>0$,
we see that the DEC does not hold at all, regardless of the sign of
$\sigma$.
\begin{figure}
\begin{center}
\epsfig{figure=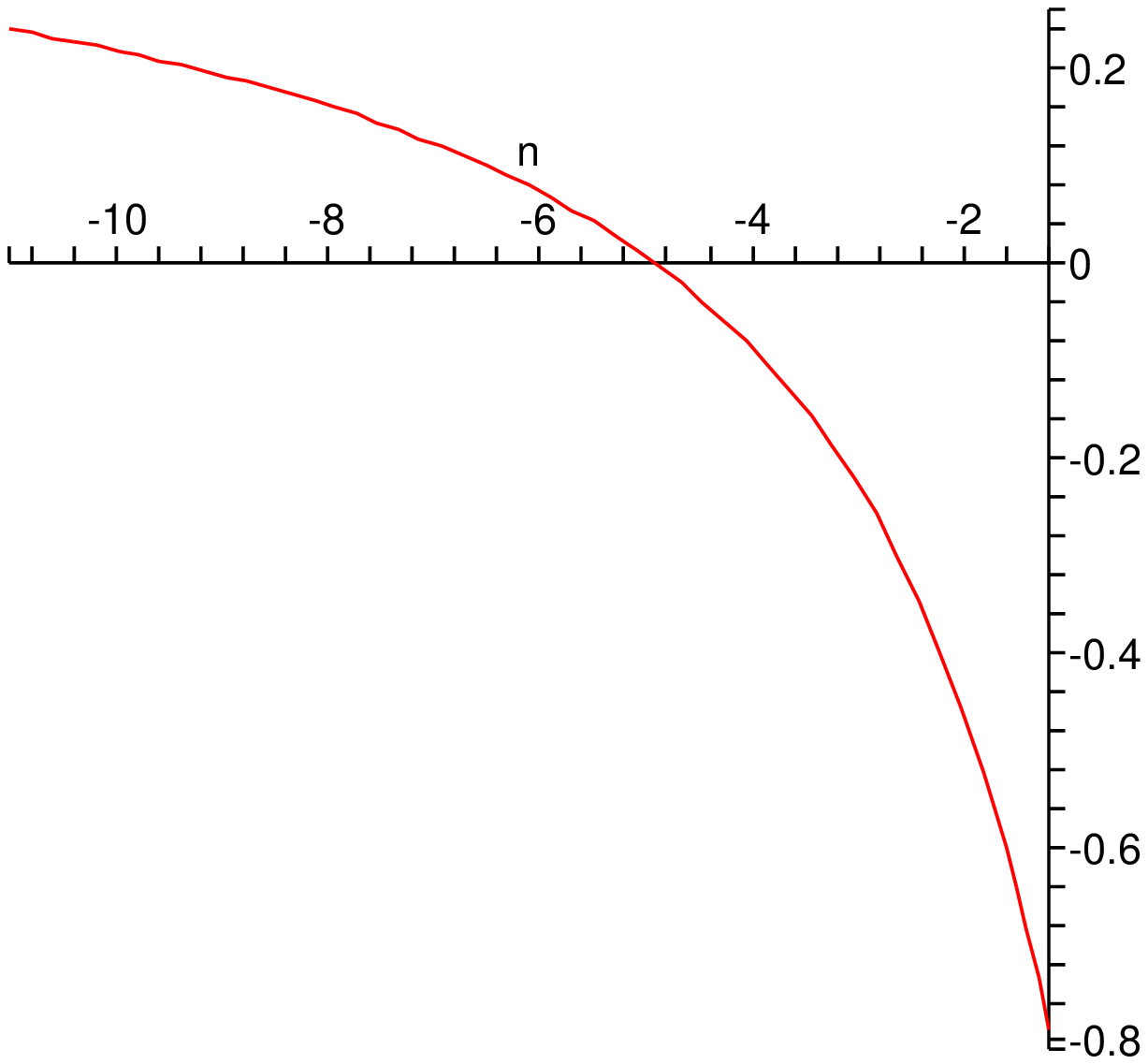,width={7cm}}\hspace{6mm}
\epsfig{figure=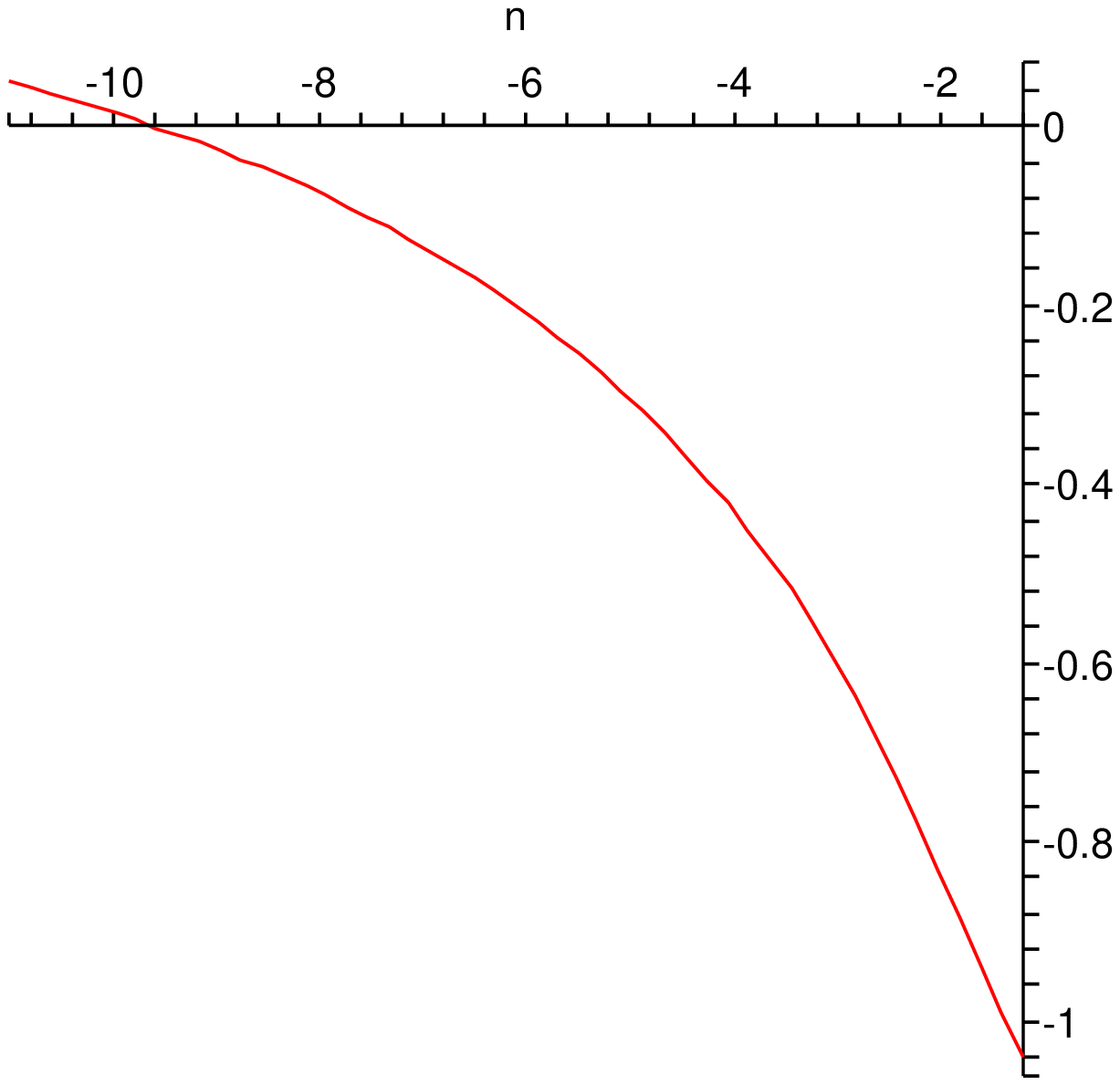,width={7cm}}
\end{center}
\caption{\footnotesize  The behavior of SEC and NEC as a function of
$n$ for $n<0$ and $\sigma>0$, left  and  of SEC and NEC for $n<0$
and $\sigma<0$, right.}
\end{figure}
\section{Conclusions}
In this work we have studied the energy conditions in the BD theory
and compared the results with that of $f(R)$ gravity, benefiting
from the ease with which the parameters of interest can be derived
in the latter and subsequently used in the former. This would help
us to check the validity of the  energy conditions in BD theory by
invoking the energy conditions derived from a generic $f(R)$ theory.
The parameters involved were shown to be consistent and compatible
with an accelerated expanding universe.

\section*{Acknowledgment}
One of us, YT, is grateful to S. Jalalzadeh for valuable
discussions.

\end{document}